\documentclass[twocolumn,useAMS,usenatbib]{mn2e}

\usepackage{amsmath}
\usepackage{amssymb}
\usepackage{graphicx}
\usepackage{mathrsfs}
\usepackage[retainorgcmds]{IEEEtrantools}
\usepackage[english]{babel}
\usepackage{upgreek}
\usepackage{color}
\usepackage{hyperref}
\DeclareMathOperator{\abs}{abs}
\DeclareMathOperator{\sign}{sign}

\newcommand{\ud}{\mathrm{d}}
\newcommand{\ug}{\mathrm{g}}
\newcommand{\uA}{\mathrm{A}}
\newcommand{\uG}{\mathrm{G}}
\newcommand{\uM}{\mathrm{M}}
\newcommand{\uN}{\mathrm{N}}
\newcommand{\uS}{\mathrm{S}}
\newcommand{\PW}{\mathrm{PW}}
\newcommand{\VTST}{\mathrm{VTST}}
\newcommand{\full}{\mathrm{full}}
\newcommand{\kin}{\mathrm{kin}}

\renewcommand{\tabcolsep}{0.17cm}

\title[MHD jets with relativistic gravity]{Linking accretion flow and particle acceleration in jets -- II. Self-similar jet models with full relativistic MHD gravitational mass}

\author[Peter Polko, David L. Meier and Sera Markoff]{Peter Polko,$^{1}$\thanks{Present address: John S. Toll Building, University of Maryland, College Park, MD 20742-4111, USA.}\thanks{E-mail: polko@umd.edu} David L. Meier$^{2}$ and Sera Markoff$^{1}$\\
$^{1}$Astronomical Institute `Anton Pannekoek', University of Amsterdam, PO Box 94249, 1090 GE Amsterdam, the Netherlands\\
$^{2}$Jet Propulsion Laboratory, California Institute of Technology, Pasadena, CA 91109, USA}

\pagerange{\pageref{firstpage}--\pageref{lastpage}} \pubyear{2012}

\begin{document}
\label{firstpage}
\maketitle

\begin{abstract}
We present a new, semi-analytic formalism to model the acceleration and collimation of relativistic jets in a gravitational potential. The gravitational energy density includes the kinetic, thermal, and electromagnetic mass contributions. The solutions are close to self-similar throughout the integration, from very close to the black hole to the region where gravity is unimportant. The field lines are tied to the conditions very close to the central object and eventually overcollimate, possibly leading to a collimation shock. This collimation shock could provide the conditions for diffusive shock acceleration, leading to the observed electron populations with a power-law energy distribution in jets.

We provide the derivation, a detailed analysis of a solution, and describe the effects the parameters have on the properties of the solutions, such as the Lorentz factor and location of the collimation shock. We also discuss the deviations from self-similarity.

By comparing the new gravity term with the gravity term obtained from a non-relativistic formalism in a previous work, we show they are equivalent in the non-relativistic limit. This equivalence shows the approach taken in that work is valid and allows us to comment on its limitations.
\end{abstract}


\section{Introduction}
Jets are important building blocks of our Universe. When the supermassive black hole in the centre of a galaxy is activated by accretion, a resulting jet can significantly alter its surroundings, by heating or displacing the ambient medium, affecting the evolution of the galaxy, or transporting angular momentum, possibly changing the spin of the black hole itself. Since the central black hole is in general too small to resolve \citep[see, e.g.,][]{1999Natur.401..891J,2011Natur.477..185H,2012Sci...338..355D}, we cannot usually observe the conditions around it directly. But as the jet is formed in this region, by observing the jet at larger distances, we can obtain this valuable information indirectly, if we have a method of linking central conditions with the jet flow further out.

The signature radiation from jets is synchrotron radiation, which provides the radio band, and sometimes dominates the infrared to X-ray bands. In the radio band, the observed spectrum is often flat or slightly inverted, which, for a compact, self-absorbed jet, is generally interpreted as the superposition of several synchrotron components from an underlying electron population with a power-law energy distribution \citep{1979ApJ...232...34B}. If there were a population of electrons closer to the black hole with, for example, a quasi-thermal energy distribution, this change would result in a break in the spectrum, corresponding to the location where the particles are accelerated to a power-law distribution for the first time. If particle acceleration is present from the base of the jets, then this break would correspond to the most compact scale of the jet. Such a break has indeed been observed in several sources, both in active galactic nuclei (AGN) \citep{1999ApJ...516..672H}, and black hole X-ray binaries (BHXRBs) \citep{2002ApJ...573L..35C,2011ApJ...740L..13G,2013MNRAS.429..815R}.

Modelling of the broad-band spectrum consistently predicts the height of this location of particle acceleration to lie in the range $\sim 10\mbox{--}1000~r_\ug$, where $r_\ug$ is the gravitational radius, with a dependence on the luminosity of the source \citep{2001A&A...372L..25M,2003A&A...397..645M,2005ApJ...635.1203M,2007ApJ...670..600G,2007ApJ...670..610M,2008ApJ...681..905M,2009MNRAS.398.1638M}. High-energy electrons would quickly lose their energy via synchrotron radiation, inverse Compton and adiabatic losses. Since the observed flat spectrum implies the power-law distribution persists for times longer than this cooling time, it seems that the particles are accelerated continuously beyond this location \citep[][but see also \citealt{2009ApJ...699.1919P}]{2001A&A...373..447J}.

A natural way to accelerate particles into a power-law distribution is via diffusive shock acceleration \citep{1949PhRv...75.1169F,1978MNRAS.182..147B,1983RPPh...46..973D,2007Ap&SS.309..119R}. A jet can be collimated by forces such as the pressure of the surrounding medium, or the magnetic tension of the field lines. If these forces are strong enough, the jet may eventually contract, or overcollimate, guiding the matter towards the jet axis. If this happens, a resulting collimation shock can provide the required energy and environment to accelerate the particles into the observed power-law distribution. Since particles seem to be accelerated continuously, this shock would have to be a stable feature in the jet. This condition can be satisfied beyond the modified fast point (MFP), a singular point in a magnetohydrodynamic (MHD) flow, since at the MFP the flow upstream is causally disconnected from the flow downstream. A shock beyond the MFP is therefore unable to disrupt the flow causing it. Around the MFP the flow also can start to overcollimate. For these reasons we use the MFP as a proxy for the start of the acceleration region, even though the latter may lie at twice the height of the MFP itself. Although previous studies stated that in a relativistic formalism the MFP could only lie at infinity \citep{1982MNRAS.199..883B,1992ApJ...394..459L,2003ApJ...596.1080V}, we showed in an earlier paper it is possible to cross the MFP at a finite height \citep[hereafter Paper I]{2010ApJ...723.1343P}.

By developing an MHD formalism describing a jet, we would like to explore the possibility that the ideas given above can explain the observed jet features. Such a jet model could equate the location of a possible collimation shock with conditions close to the black hole where the jet is formed. In order to be able to have a reliable model close to the black hole, the flow needs to first cross two other singular points, the Alfv\'en point, where the flow towards the axis matches the Alfv\'en velocity, and the modified slow point (MSP), where it matches the magnetosonic slow velocity. The MSP cannot form without gravity. Since in a relativistic self-similar MHD framework gravity scales differently than the other physical quantities, for a purely self-similar model gravity needs to be neglected. By comparing a relativistic formalism with a non-relativistic one including gravity, in a previous paper we showed it is possible to include gravity into the relativistic equations and give a prescription for the region where self-similarity holds to within a specified tolerance \citep[hereafter Paper II]{2013MNRAS.428..587P}.

In this paper we derive the gravity term from the general MHD equations. We show that the previously derived gravity term is the part representing the kinetic mass contribution. We will therefore call it the `kinetic' gravity term for brevity. With this comparison, we demonstrate that the previous approach is valid and we also can explore its limitations better. The new model presented here, is valid in a wider range of cases, since it also takes the thermal and electromagnetic mass contributions into account. We will therefore call the newly obtained term the `full' gravity term for contrast. The new model also keeps track of the fraction of the total energy which is present in the gravitational field.

The observed Lorentz factors of jets in AGN are $\lesssim 10$ \citep{2009AJ....138.1874L}, while for BHXRBs they are found to be $\lesssim 2$ \citep{1999ARA&A..37..409M}. We will keep these Lorentz factors and MFP heights in mind when we explore the parameter space of the solutions in this new model.

In Section \ref{sec:fullmethod}, we describe the derivation of the full gravity term, the changes made to the previous model, and the effects these changes have on the solutions. In Section \ref{sec:fullresults}, we give an overview of the solutions we find and a preliminary parameter study determining the range of properties. In Section \ref{sec:fulldiscussion}, we discuss our findings and present our conclusions.

\section{Method}
\label{sec:fullmethod}
In this section, we briefly review the preceding work, give the derivation of the full gravity term including the contributions to the mass, list the modifications to the equations, and describe the approach to finding solutions.

\subsection{Background}
In Paper II we showed that it is possible to include gravity in a single relativistic wind equation, which describes the acceleration along a field line, by comparison of a relativistic formalism without gravity \citep{2003ApJ...596.1080V} with a non-relativistic formalism including gravity \citep[hereafter VTST00]{2000MNRAS.318..417V}. By making a term-by-term comparison of the wind equations of the two formalisms, we were able to isolate the term equating to gravity and derive a single wind equation relating the region where gravity is important and the flow has non-relativistic velocities, to the region where gravity can be neglected and the flow is relativistic.

While this approach gives solutions that satisfy self-similarity to some degree, it still has several stringent constraints. In special relativistic MHD a characteristic velocity appears, the velocity of light, as well as a corresponding length scale, the light cylinder radius. The velocity of light has no radial dependence, which means gravity cannot be included in a self-similar way \citep{1992ApJ...394..459L}. Therefore, only when the flow has non-relativistic velocities at the base of the jet where gravity is important, are the equations self-similar with gravity. This condition implies either that the gravitational potential is not very strong because the MSP occurs relatively far away from the black hole, or that the temperature or magnetic field energy density at the base of the flow is low, so the initial thermal or magnetic acceleration is not enough to accelerate the matter to relativistic velocities. Another limitation is the gravity term itself. Since this term is derived from a non-special-relativistic formalism, it only includes the kinetic mass contribution, while the thermal and electromagnetic mass contributions are ignored. Since even the kinetic part misses a relativistic correction (see \S \ref{subsec:gravitytermcomparison}), in relativistic flows the kinetic gravity term is only a poor approximation of the full gravitational effects.

For these reasons, we aim to derive a gravity term taking into account all mass contributions, and relativistic effects. The approach we use, similar to the approach taken in Paper I and II, is to derive the wind equation (the equation where the singular points manifest themselves) from the energy equation (which describes the forces along a field line, equation \eqref{eq:energyequation}) and the transfield equation (which describes the forces across a field line). We rewrite the derivative of the energy equation with respect to $\theta$, the poloidal spherical angle that we use as our independent variable, and the transfield equation in the general form:
\begin{equation}
A_1 \frac{\ud M^2}{\ud \theta} + B_1 \frac{\ud \psi}{\ud \theta} = C_1 \mbox{ and } A_2 \frac{\ud M^2}{\ud \theta} + B_2 \frac{\ud \psi}{\ud \theta} = C_2,
\label{eq:determinant}
\end{equation}
where $M$ is the Alfv\'enic Mach number, $\psi$ is the angle a field line makes with the disc, $A$, $B$, and $C$ are functions not containing these derivatives of $M$ and $\psi$, and where subscript 1 refers to functions from the energy equation, and subscript 2 refers to functions from the transfield equation. From these two equations we can solve for the wind equation (given by $\ud M^2/\ud \theta$) \emph{without} gravity by using the determinant method:
\begin{equation}
\frac{\ud M^2}{\ud \theta} = \frac{C_1 B_2 - C2 B_1}{A_1 B_1 - A_2 B_1}.
\label{eq:windequation}
\end{equation}
For the full equations see Appendix \ref{app:equations}.

\subsection{The new gravity term}
\label{subsec:gravityterm}
In order to derive a fully relativistic gravity term with all mass contributions, we start with the fully special relativistic equations. Making the assumptions of time independence, axisymmetry and self-similarity, the energy equation along each magnetic field line gives a conserved quantity, the Bernoulli constant or specific total energy:
\begin{equation}
(\mu - 1) c^2 + \mu \Phi = \mathrm{constant},
\end{equation}
where $\mu c^2$ is the total energy-to-mass flux ratio, or specific internal energy of the plasma, including rest mass, $c$ is the velocity of light and $\Phi$ is the gravitational potential \citep{2012bhae.book.....M}. If gravity is neglected, $\mu = \mathrm{constant}$, since $\Phi = 0$, but if gravity is taken into account $\mu$ becomes a variable. For ease of notation, we define the Bernoulli constant as
\begin{equation}
\mu' \equiv \mu + \mu \frac{\Phi}{c^2} = \mu \left(1 + \frac{\Phi}{c^2}\right),
\label{eq:muprimepot}
\end{equation}
thus far away from the black hole, where gravity is unimportant, $\mu$ approaches this constant. The gravitational potential in the Newtonian limit is given by
\begin{equation}
\frac{\Phi_\uN}{c^2} = - \frac{r_\ug}{r} = - \frac{\mathcal{G M}}{c^2} \frac{\sin(\theta)}{\varpi_\uA G},
\label{eq:gravitationalpotential}
\end{equation}
where $r_\ug$ is the gravitational radius, $r$ the spherical radius, $\mathcal{G}$ is the gravitational constant, $\mathcal{M}$ is the mass of the black hole, $\varpi_\uA$ is the cylindrical radius of the Alfv\'en point and $G$ is the cylindrical radius in units of $\varpi_\uA$. For conciseness, we will use the expression for the Newtonian limit in the remainder of the equations. For our calculations, we used the pseudo-Newtonian Paczy\'nsky--Wiita potential \citep{1980A&A....88...23P}:
\begin{equation}
\frac{\Phi_\PW}{c^2} = - \frac{r_\ug}{r - r_\uS} = - \left[ \frac{c^2 \varpi_\uA G}{\mathcal{G M} \sin(\theta)} - 2 \right]^{-1},
\label{eq:PaczynskyWiita}
\end{equation}
where $r_\uS$ is the Schwarzschild radius, equal to twice the gravitational radius. To obtain the equations with the Paczy\'nsky--Wiita potential, replace wherever the right-hand side of equation \eqref{eq:gravitationalpotential} appears with the right-hand side of equation \eqref{eq:PaczynskyWiita}. Although in this section the gravitational radius, $\mathcal{G M}/c^2$ is still explicit, in the results below we absorb it into $\varpi_\uA$ so this distance is expressed in gravitational radii. A subscript A denotes a value at the Alfv\'en point. Using equations \eqref{eq:muprimepot} and \eqref{eq:gravitationalpotential} we can calculate $\mu'$ at the Alfv\'en point and consequently $\mu$ at every point along the field line using equation \eqref{eq:muprimepot}.

If we substitute \eqref{eq:muprimepot} into the energy equation \eqref{eq:energyequation} and derive the resulting equation with respect to $\theta$, we obtain an additional term to $C_1$, which we denote by $C_1^+$ and is given to first order by
\begin{equation}
C_1^+ = - \frac{\mathcal{G M}}{c^2} \frac{\sin(\theta)}{\varpi_\uA G} \left[ \frac{\xi^2 x_\uA^4}{F^2 \sigma_\uM^2} \frac{\cos^2(\psi + \theta)}{\sin^2(\theta)} + \frac{M^4}{G^4} \right],
\end{equation}
where $\xi c^2$ is the specific relativistic enthalpy, $x$ is the cylindrical radius in units of the light cylinder radius ($x = x_\uA G$), and $\sigma_\uM$ is the magnetisation parameter. $F$ is the parameter that controls the current distribution and is given by $F = 1 + \ud \log I / \ud \log \varpi$. If $F > 1$, the current increases with radius and this case is called the current-carrying regime. The case $F < 1$ is called the return-current regime, which is the case for our solutions since we set $F = 0.75$. In this regime, the electrical force acts to collimate the jet, while the magnetic force acts to decollimate it. The gravitational force in the transfield equation is given by
\begin{equation}
f_\uG = \left( \gamma \rho_0 + \frac{\mathfrak{E}}{c^2} \right) \left( \nabla \Phi \cdot \hat{\boldsymbol{n}} \right),
\end{equation}
where $\gamma$ is the Lorentz factor, $\rho_0$ is the matter density, $\hat{\boldsymbol{n}}$ is the unit vector in the transfield direction and $\mathfrak{E}$ is the energy density, which under the assumption of flat space--time is given by
\begin{equation}
\mathfrak{E} = \gamma (\gamma - 1) \rho_0 c^2 + \gamma^2 \frac{\Gamma}{\Gamma - 1} P - P + \frac{1}{8 \uppi} \left(B^2 + E^2 \right),
\end{equation}
where $\Gamma$ is the polytropic index, $P$ is the pressure, $B$ is the magnetic field, and $E$ is the electric field \citep{2012bhae.book.....M}. If we fill in the expressions for these quantities and use the scaling for the transfield equation so it has the same `units' as the energy equation, we obtain the addition to $C_2$:
\begin{IEEEeqnarray}{rCl}
C_2^+ & = & \Bigg\{ \frac{\mu^2 x^4}{F^2 \sigma_\uM^2} \left[ \frac{1}{M^2} \frac{(1 - M^2 - x_\uA^2)^2}{(1 - M^2 - x^2)^2} + \frac{x^2 (1 - G^2)^2}{2 G^4 (1 - M^2 - x^2)^2} \right] \IEEEnonumber \\
& & - \frac{x^4}{F^2 \sigma_\uM^2} \frac{\Gamma - 1}{\Gamma} \frac{\xi(\xi - 1)}{M^2} + \frac{1}{2} \frac{(1 + x^2) \sin^2(\theta)}{\cos^2(\psi + \theta)} \Bigg\} \IEEEnonumber \\
& & \times \left[ \frac{\mathcal{G M}}{c^2} \frac{\sin(\theta)}{\varpi_\uA G} \cos^2(\psi + \theta) \right].
\end{IEEEeqnarray}
The wind equation \emph{with} gravity is given by:
\begin{IEEEeqnarray}{rCl}
\frac{\ud M^2}{\ud \theta} & = & \frac{(C_1 + C_1^+) B_2 - (C_2 + C_2^+) B_1}{A_1 B_2 - A_2 B_1} \IEEEnonumber \\
& = & \frac{C_1 B_2 - C_2 B_1 + \mathscr{G}_\full}{A_1 B_2 - A_2 B_1},
\end{IEEEeqnarray}
so the gravity term including all mass contributions, which we will call the `full' gravity term, $\mathscr{G}_\full$, added to the numerator of the wind equation \emph{without} gravity is then given by $C_1^+ B_2 - C_2^+ B_1$, where:
\begin{equation}
B_1 = \frac{M^4}{G^4},
\end{equation}
and:
\begin{equation}
B_2 = \left[ \frac{(1 - x^2)}{\cos^2(\psi + \theta)} - M^2 \right] \sin^2(\theta),
\end{equation}
are the scaled components from the energy and transfield equation, respectively. Putting all these parts together gives us:
\begin{IEEEeqnarray}{rCl}
\mathscr{G}_\full & = & \textcolor{red}{- \frac{\mathcal{G M}}{c^2} \frac{\sin(\theta)}{\varpi_\uA G}} \Bigg\{ \textcolor{red}{\frac{\mu^2 x_\uA^4}{F^2 \sigma_\uM^2} \frac{(1 - M^2 - x_\uA^2)^2}{(1 - M^2 - x^2)^2}} \left( 1 - x^2 \right) \IEEEnonumber \\
& & - \frac{\mu^2 x_\uA^4}{F^2 \sigma_\uM^2} \frac{x^2 (G^2 - M^2 - x^2)^2}{G^4 (1 - M^2 - x^2)^2} \left( 1 - x^2 \right) \IEEEnonumber \\
& & + \frac{\mu^2 x_\uA^4}{F^2 \sigma_\uM^2} \frac{M^2 x^2}{G^4} \bigg[ \frac{(G^2 - M^2 - x^2)^2}{(1 - M^2 - x^2)^2} \IEEEnonumber \\
& & + \frac{1}{2} \frac{M^2 (1 - G^2)^2}{(1 - M^2 - x^2)^2} \bigg] \cos^2(\psi + \theta) \IEEEnonumber \\
& & - \frac{x_\uA^4}{F^2 \sigma_\uM^2} \frac{\Gamma - 1}{\Gamma} \xi (\xi - 1) M^2 \cos^2(\psi + \theta) \IEEEnonumber \\
& & + \frac{1}{2} \frac{M^4}{G^4} (1 + x^2) \sin^2(\theta) \Bigg\}.
\label{eq:fullgravity}
\end{IEEEeqnarray}

\subsection{Comparison with the kinetic gravity term}
\label{subsec:gravitytermcomparison}
When we compare the gravity term obtained in Paper II:
\begin{equation}
\mathscr{G}_\kin = \textcolor{red}{- \frac{\mathcal{G M}}{c^2} \frac{\sin(\theta)}{\varpi_\uA G} \frac{\mu^2 x_\uA^4}{F^2 \sigma_\uM^2} \frac{(1 - M^2 - x_\uA^2)^2}{(1 - M^2 - x^2)^2}},
\end{equation}
with the full gravity term given above, we can see the kinetic gravity term corresponds to the first line of equation \eqref{eq:fullgravity}, coloured red, not including the factor $(1 - x^2)$. For small $x$ this factor approaches unity. The first line of equation \eqref{eq:fullgravity} corresponds to the kinetic energy contribution to $\mathfrak{E}$, which is why we will refer to the previous gravity term derived in Paper II as the kinetic gravity term. The other lines correspond to the thermal, electric and magnetic energies. If these energies are unimportant and $x$ is small (i.e., the field line lies well within the light cylinder radius) as is the case in VTST00, the two gravity terms are similar in value. For solutions with $x$ close to 1, the kinetic gravity term is a poor approximation, as it will always overestimate the strength of gravity.

We can label individual field lines with the dimensionless variable $\alpha$, defined as the square of the cylindrical radius of the Alfv\'en point over a reference cylindrical radius equal for all field lines, $\alpha \equiv \varpi_\uA^2 / \varpi_0^2$. Since $\varpi_0$ is a constant, acting as a unit length and introduced purely to define a dimensionless distance, $\varpi_\uA \propto \alpha^{1/2}$. Because we express $\varpi_\uA$ in gravitational radii, we can also express $\varpi_0$ in gravitational radii and let the mass of the black hole determine the physical distance. Since the dimensionless variables are scaled back with this same reference length to obtain the physical values, we can freely choose the reference length and label a particular field line with $\alpha = 1$.

All the physical variables, including the gravity term through $\varpi_\uA$, scale with $\alpha$, which for a fully self-similar solution is equivalent to scaling with spherical radius. The scaling with $\alpha$ is the same for both the kinetic and the full gravity term, namely $\alpha^{-1/2}$. A different scaling with $\alpha$ means a different dependence on the radius, and therefore a different self-similarity. Since the full gravity term has a different dependence on $\alpha$ from the rest of the wind equation ($\alpha^0$), it violates the self-similarity assumption the same way the kinetic gravity term does in Paper II. For this reason we postulate that the field lines controlling the location of the MFP originate in a small region of the accretion disc and can be approximated by a flux tube. The width of this flux tube then determines the deviation from self-similarity. We therefore argue that our solutions are valid within a flux tube of a certain width, as laid out in Paper II. Most of the activity of jets is concentrated in specific regions. The emission in the case of the compact jet cores most relevant for a steady-state treatment, for example, predominantly arises in the photosphere of the jet, a layer of limited depth. We therefore think that a flux tube provides enough information to reproduce most of the features of observed jets.

\subsection{Effects of the full gravity term}
There are several changes we have to make to the equations in order to be self-consistent when using the full gravity term. Due to the fact that $\mu$ is no longer a constant along a field line, we have to calculate it at every step using equation \eqref{eq:muprimepot} from $\mu'$ calculated at the Alfv\'en point. Since the potential is negative, $\mu$ decreases outwards, eventually asymptotically approaching $\mu'$. Going towards the black hole, $\mu$ can increase without limit. Since the Lorentz factor of the flow is proportional to $\mu$, it is possible for the Lorentz factor to decrease when the flow moves outwards. This decrease may sound counterintuitive since the flow is accelerating, but this acceleration is in the poloidal plane, while the Lorentz factor also includes the velocity in the azimuthal direction. The decrease of the Lorentz factor is thus interpreted as the initial relativistic velocities in the azimuthal direction gradually changing into a mainly poloidal flow.

\subsection{Approach to finding solutions}
\label{subsec:approach}
By including gravity we have introduced a new characteristic length, the gravitational radius. The parameter $\varpi_\uA$ is the cylindrical radius of the Alfv\'en point. Since only the ratio of $\varpi_\uA$ and $\mathcal{M}$ appears in the equations, reflecting mass scaling, it leads to no loss of generality if we express the Alfv\'en radius in gravitational radii, and incorporate the factor $\mathcal{G M}/c^2$ into $\varpi_\uA$. This is the convention we will use in the remainder of the paper. Solutions are therefore no longer directly dependent on the mass of the black hole, but it does still set the overall scale of the system.

\begin{table*}
\begin{center}
\begin{tabular}{l c c c c c c c c c c c c c c}
\hline
& & & $\theta_\uA$ & $\psi_\uA$ & & $\varpi_\uA$ & $\mathcal{M}$ \\
& $F$ & $\Gamma$ & (${}^\circ$) & (${}^\circ$) & $\sigma_\uM$ & ($r_\ug$) & (M$_\odot$) & \multicolumn{2}{c}{$x_\uA^2$} & \multicolumn{2}{c}{$q$} & $p_\uA$ & $\sigma_\uA$ & $\mu'$ \\
\hline
First solution & 0.75 & 5/3 & 60 & 45 & 0.000\,785\,798 & 18.2088 & 10 & \multicolumn{2}{c}{0.01} & \multicolumn{2}{c}{0.014\,359} & $-5.759\,03$ & 0.010\,956 & 1.069\,56 \\
Reference & 0.75 & 5/3 & 60 & 40 & 0.02 & 15 & 10 & \multicolumn{2}{c}{0.145\,330} & \multicolumn{2}{c}{0.024\,184} & $-7.946\,40$ & 0.186\,323 & 1.313\,44 \\
\hline
\\
\end{tabular}\\
\end{center}
\caption{
Parameters of the first solution and the reference solution used in the parameter study. $p_\uA$ is the value of $\ud M^2 / \ud \theta$ at the Alfv\'en point, and $\sigma_\uA$ the magnetisation function evaluated at the Alfv\'en point. The values for the first seven parameters ($F$ through $\mathcal{M}$) of the solution in this work are exact, except $\sigma_\uM$ for the first solution, for the last five ($x_\uA^2$ through $\mu'$) they are rounded off, except $x_\uA^2$ for the first solution. Because singular solutions require high precision, the rounded-off numbers are given with six significant digits. For this initial solution, $\sigma_\uM$ was used as a fitting parameter instead of $x_\uA^2$, since we wanted to ensure $x_\uA^2$ was small.
}
\label{tab:fullgravityparametertable}
\end{table*}

\begin{figure}
\begin{center}
\includegraphics[width = 0.45 \textwidth]{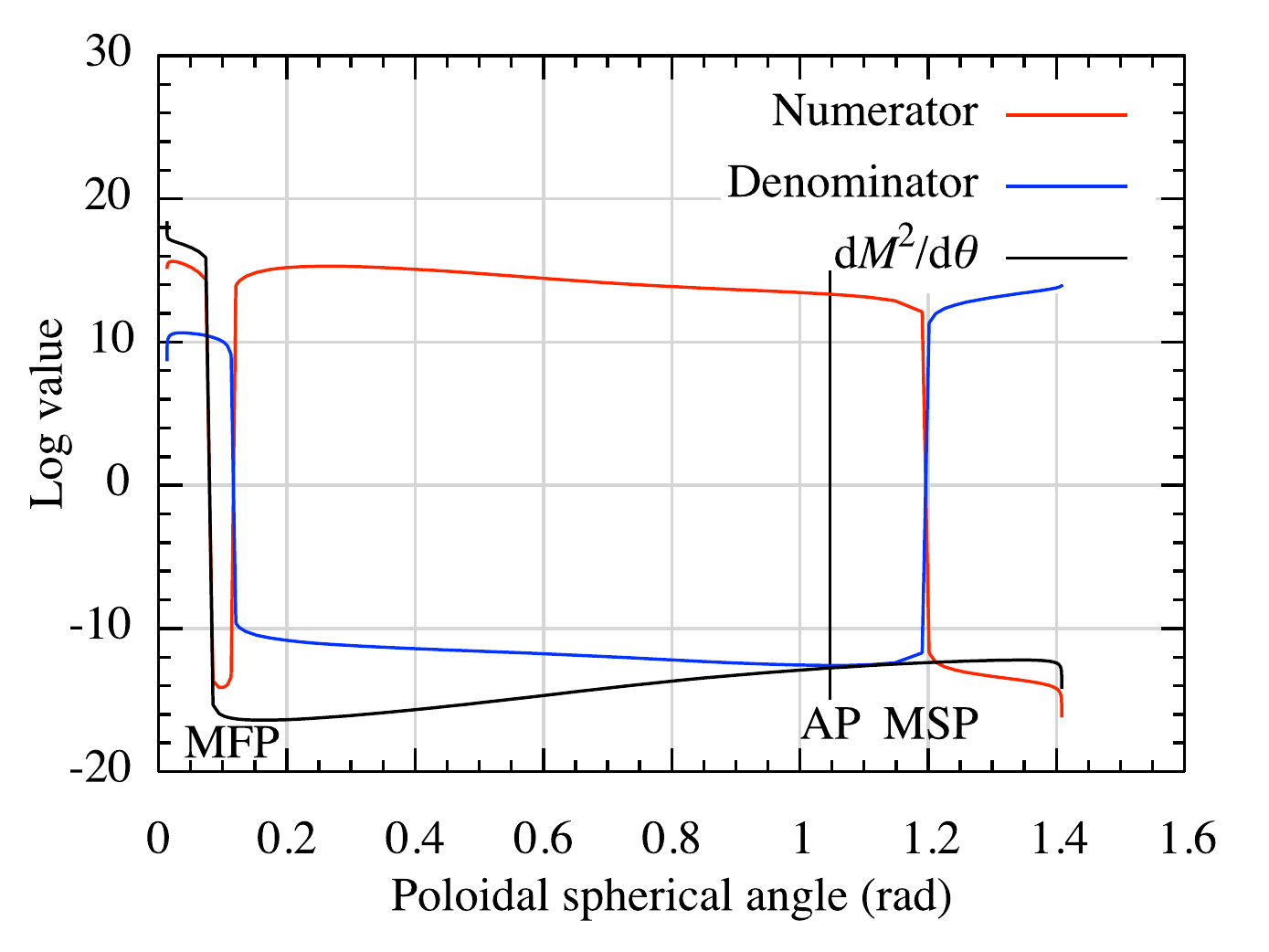}
\end{center}
\caption{
First solution found using the full gravity term showing both an MFP and MSP\@. The red and blue lines show the numerator and denominator of the wind equation, respectively, while the black line shows their ratio. The latter determines the total bulk acceleration of $M^2$ with respect to polar angle $\theta$. The vertical line shows the location of the Alfv\'en point (AP). The vertical axis is a `scaled logarithm' of the plotted parameters, i.e., $y = \sign(x) \log_{10} [1 + \abs(x)/10^{-12}]$ to clearly show the variables over many orders of magnitude. The integration is continued beyond the MFP and below the MSP to show $\ud M^2 / \ud \theta$ is smooth, but since at the singular points the ratio is almost $0 / 0$, we do not consider these regions for further analysis. Note that since the angle $\theta$ decreases with increasing height a negative value of $\ud M^2 / \ud \theta$ corresponds to acceleration.}
\label{fig:fullnumden}
\end{figure}

As mentioned above, the gravity term has a different $\alpha$ dependence than the other terms in the numerator of the wind equation. While the simple scaling of true self-similarity is therefore no longer possible, we have found that by changing $\varpi_\uA$ we can populate our jet with field lines and mimic this scaling to a high degree (see Section \ref{subsec:fullselfsimilarity}). By using the Alfv\'en regularity condition we ensure a smooth crossing at the Alfv\'en point. Since we want our solutions to cross also both the MFP and MSP, and these crossings correspond to one boundary condition each, we need to fit two parameters to ensure a smooth crossing at those locations as well. We have chosen to use $x_A^2$ and $q$ for these fitting parameters, as this combination makes sure field lines do not cross when we populate the jet by changing the free parameter $\varpi_\uA$, which acts as a scaling parameter. $q$ is related to the adiabat $Q = P/\rho_0^\Gamma$, by
\begin{equation}
q = \frac{B_0 \alpha^{F - 2} x_\uA^4}{4 \uppi c^2 F^2 \sigma_\uM^2} \left( \frac{\Gamma}{\Gamma - 1} \frac{Q}{c^2} \right)^{1/(\Gamma - 1)},
\end{equation}
where $B_0$ is a reference field strength to obtain dimensionless variables in the same way as $\varpi_0$ is used.

For a new solution, we set the free parameters, $\theta_\uA$, $\psi_\uA$, $\sigma_\uM$ and $\varpi_\uA$, and then keep adjusting, for example, $x_\uA$ until the MFP has been crossed. Then we change $q$ until the integration crosses the MSP\@. This integration usually does not cross the MFP anymore, so we change $x_\uA$ again and repeat. In this way, we converge to a solution that crosses both the MFP and MSP\@.

A key issue in this approach is finding an initial solution. Once obtained, it is possible to find additional solutions by making small steps in the free parameters, avoiding any regions where solutions do not exist. Because of the diminishing effect of the $(1 - x^2)$ factor, we found there were no solutions around the parameter values of the solution described in Paper II, which had a high value of $x_\uA^2 \approx 0.75$. Since all equations, including the gravity term, revert to their non-relativistic form in the appropriate limit, we have adjusted the parameters of the solution given in fig.~4 of VTST00 and indeed found a solution nearby (see Table \ref{tab:fullgravityparametertable}). See Section \ref{subsec:firstsolution} for further details.

We found that in general the required precision to obtain a solution where the numerator and denominator of the wind equation crossed zero within the same integration step exceeds quadruple precision and involved prohibitive convergence times. For these reasons, when the denominator approached zero first, effectively ending the integration, we extrapolated the numerator and denominator and treated as a solution those integrations where the relative difference between the angle where numerator and denominator crossed zero was smaller than $10^{-4}$~radians.

\section{Results}
\label{sec:fullresults}
We will discuss the first solution found using the full gravity term. Then we present an initial exploration of parameter space with a view to finding solutions with properties that correspond to observed systems. We also detail the effects the parameters have on the solution.

\begin{figure*}
\begin{center}
\includegraphics[width = \textwidth]{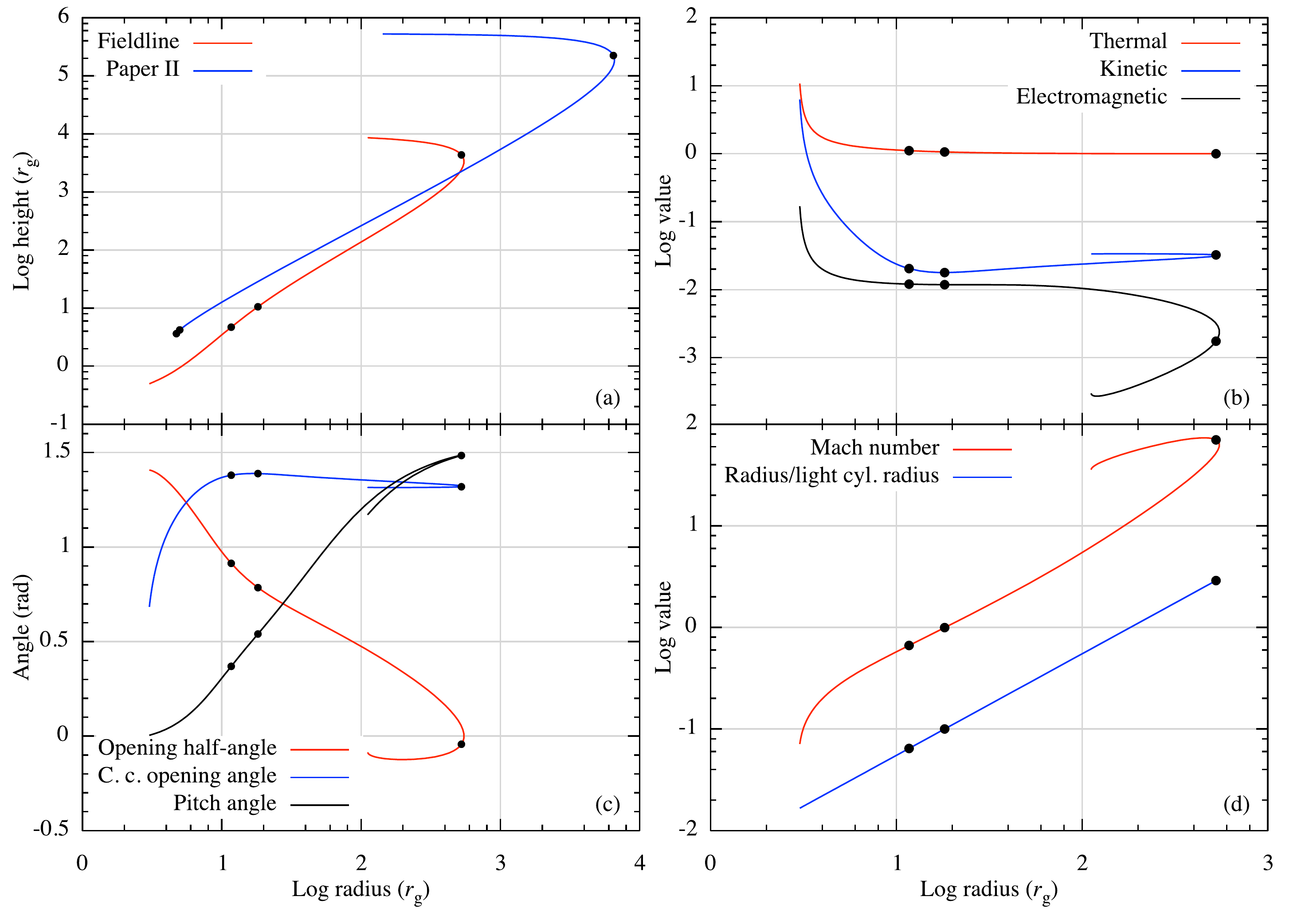}
\end{center}
\caption{
Various physical quantities that result from integrating the wind equation depicted in Fig.~\ref{fig:fullnumden}, now plotted against cylindrical radius $\varpi$ instead of $\theta$. These plots are similar to those in fig.~4 of Paper I. Panel (a) shows the geometry of the field line of this paper in red, and the field line from Paper II in blue. Panel (b) shows the magnetic energy ($S \equiv - \varpi \Omega B_{\phi} / \Psi_A c^2$), the thermal energy including the rest mass ($\xi$), and the kinetic energy [$(\gamma - 1) \xi$]. Panel (c) shows the opening half-angle of the outflow ($\uppi / 2 - \psi$), the causal connection opening angle ($\arcsin [1 / \gamma]$), and the pitch angle of the field line. Panel (d) shows the cylindrical radius in units of the `light cylinder' radius ($x$), and the Alfv\'enic Mach number ($M$). The MSP, Alfv\'en point, and MFP (from left to right) are indicated for each quantity by black dots.
}
\label{fig:fulloverview}
\end{figure*}

\subsection{First solution}
\label{subsec:firstsolution}
Due to the weakening of the gravity term by the $(1 - x^2)$ factor, we were unable to find solutions with both an MFP and MSP with parameters similar to those used in Paper II. Since our equations, including the full gravity term, reduce to the equations of VTST00, by using the parameters of a solution that crossed all three singular points in VTST00, we would be close to the values that produce a solution using the relativistic equations, even though this solution would be non-relativistic itself. We converted the parameters of the solution given in the caption of fig.~4 in VTST00 to the parameters used in our model. Most parameters are identical, we set $x_\uA^2 = 0.01$ to obtain a solution in the non-force-free limit of VTST00, and for $\varpi_\uA$ we used the definition of the mass-loss parameter, $\kappa_\VTST$:
\begin{IEEEeqnarray}{rCl}
\kappa_\VTST & = & \sqrt{\frac{\mathcal{G M}}{\varpi_* V_*^2}} = \sqrt{\frac{\mathcal{G M}}{c^2 \varpi_\uA} \frac{\mu^2 x_\uA^4}{F^2 \sigma_\uM^2} \frac{(1 - M^2 - x_\uA^2)^2}{(1 - M^2 - x^2)^2}},\IEEEeqnarraynumspace
\end{IEEEeqnarray}
evaluated at the Alfv\'en point:
\begin{equation}
\varpi_\uA = \frac{\lambda_\VTST^2}{\kappa_\VTST^2 x_\uA^2} \left( \frac{2 x_\uA^2 \cos(\psi_\uA)}{p_\uA \sin(\theta_\uA) \cos(\theta_\uA + \psi_\uA)}+1 \right)^{-2},
\end{equation}
where $\lambda_\VTST$ is the specific angular momentum as defined in VTST00 and we used the equations:
\begin{IEEEeqnarray}{rCl}
\left( \frac{1 - M^2 - x_\uA^2}{1 - M^2 - x^2} \right)_\uA & = & \frac{1}{\sigma_\uA + 1}, \\
\sigma_\uA & = & \frac{2 x_\uA^2 \cos(\psi_\uA)}{p_\uA \sin(\theta_\uA) \cos(\theta_\uA + \psi_\uA)}, \\
\lambda_\VTST^2 & = & \frac{\mu^2 x_\uA^6}{F^2 \sigma_\uM^2}.
\end{IEEEeqnarray}
Using the remaining $\sigma_\uM$ and $q$ as fitting parameters, we indeed obtained a solution with parameters given in Table \ref{tab:fullgravityparametertable}.

The solution found crosses all three singular points as can be seen in Fig.~\ref{fig:fullnumden}. The MFP is located at $\theta = 6^{\circ}.9$, the Alfv\'en point at $\theta = 60^{\circ}$ and the MSP at $\theta = 68^{\circ}$.

\begin{figure*}
\begin{center}
\includegraphics[width = \textwidth]{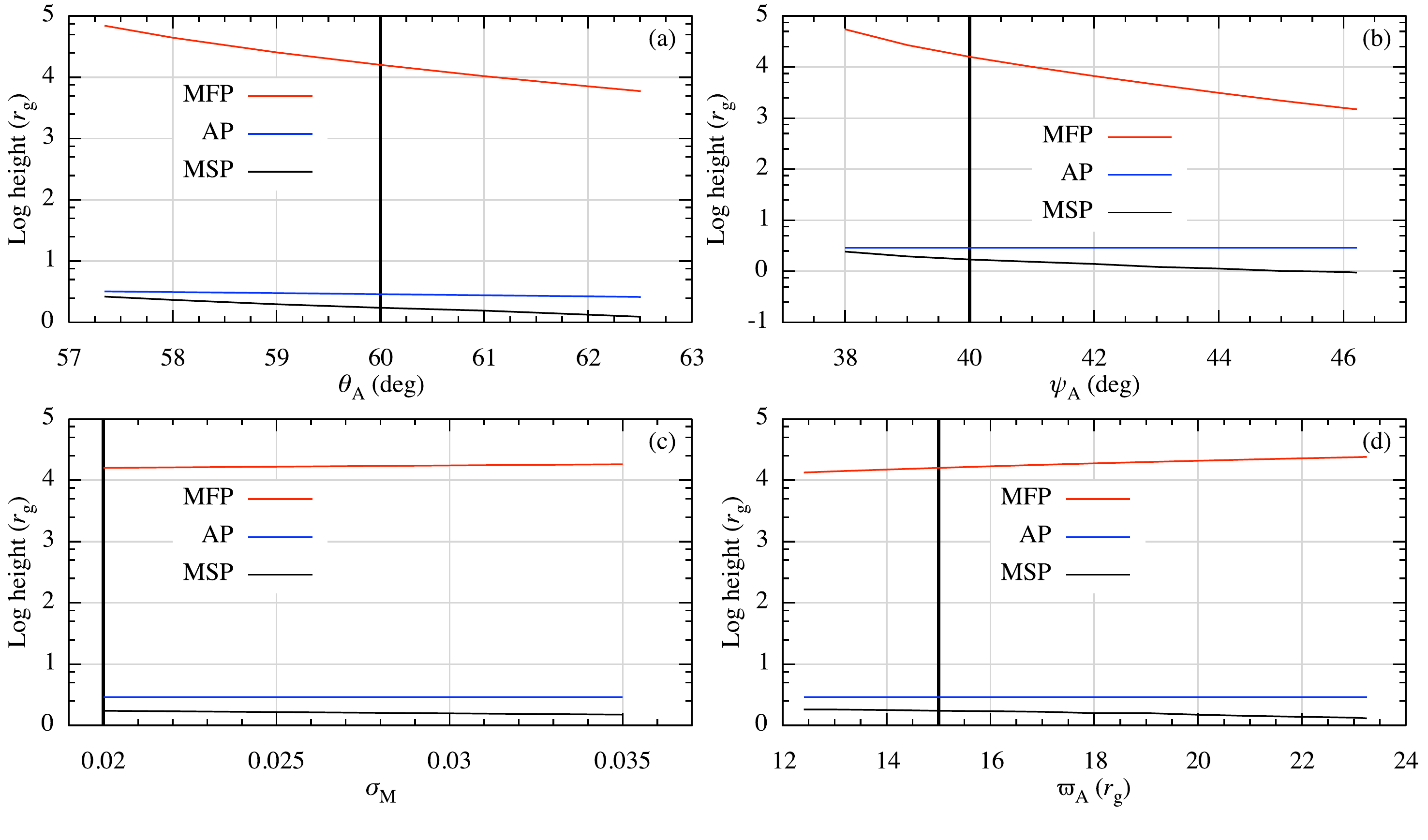}
\end{center}
\caption{
Heights (in gravitational radii) of the solutions' three singular points as a function of the four principal free parameters. The red line shows the MFP, the blue line the Alfv\'en point (AP), and the black line the MSP\@. The reference solution (see Table \ref{tab:fullgravityparametertable}) is indicated by the vertical black line. The parameters varied are as follows: panel (a): poloidal spherical angle of the Alfv\'en point ($\theta_\uA$); panel (b): the angle the field line makes with the disc at the Alfv\'en point ($\psi_\uA$); panel (c): magnetisation ($\sigma_\uM$) and panel (d): cylindrical radius of the Alfv\'en point in $r_\ug$. The horizontal axis in each of these plots is linear, not logarithmic. Note that the MSP and Alfv\'en point heights are still outside the black hole horizon (see Fig.~\ref{fig:fullalfvendistance}); i.e., even though the height of the MSP is smaller than the Schwarzschild radius ($z_\mathrm{MSP} < 2~r_\ug$), the spherical radius of the MSP for these solutions remains outside the black hole horizon ($r_\mathrm{MSP} > 2~r_\ug$).
}
\label{fig:fullMFPMSPlocation}
\end{figure*}

Panel (a) of Fig.~\ref{fig:fulloverview} shows that the field line geometry is mostly parabolic, with $\log(z)/\log(r) \approx 1.5$. In comparison with the solution found in Paper II shown in blue, the slope is slightly steeper and the MSP and Alfv\'en point lie farther away from the black hole, while the MFP lies significantly closer at a height of $4298~r_\ug$.

Panel (b) shows the thermal ($\xi$), kinetic ($(\gamma - 1) \xi$) and electromagnetic energy ($S$). Since the poloidal velocity is always at least 10 times as high as the toroidal velocity, the total kinetic energy and the poloidal kinetic energy (not plotted) are indistinguishable. The thermal energy drops monotonically to very close to 1, corresponding to thermal acceleration and cooling through adiabatic expansion, and leaving the matter in the jet cold. The electromagnetic energy also drops, which means the jet is also magnetically accelerated. Despite these modes of acceleration the Lorentz factor starts out very high and initially drops, before eventually increasing again. This behaviour is caused by the gravitational potential. As the matter climbs out of the potential well, it decelerates, despite the conversion of thermal and magnetic energy. See Section \ref{sec:fulldiscussion} for a further discussion. The Lorentz factor reaches a value of 1.07 at the MFP, showing this solution is indeed non-relativistic. Due to the low value of the magnetisation parameter, most of the energy at this point is provided by the rest mass of the matter, and the result is a kinetically dominated jet.

Panel (c) shows the opening half-angle, $\uppi / 2 - \psi$, the causal connection opening angle, $\arcsin(1/\gamma)$, and the pitch angle of the magnetic field line, $\arctan(B_\phi/B_p)$. The opening half-angle shows that the jet starts out very wide and slowly collimates, until it overcollimates around the MFP\@. In contrast the solution in Paper II started out nearly vertical, widened first and only then started to collimate. The causal connection opening angle is the relativistic equivalent of the angle of a Mach cone for supersonic velocities, describing the cone in which material can be affected by a certain location in the jet. Initially this angle is very small due to the high Lorentz factor and then increases as the flow slows down, the opposite behaviour as in Paper II. The pitch angle shows how toroidal the field lines are. If the angle is $0^{\circ}$ the field lines are purely poloidal, if $90^{\circ}$ they are purely toroidal. The flow starts mostly poloidal, becomes almost purely toroidal at the point of recollimation, and then returns to a more poloidal configuration. Beyond the light cylinder radius 80\% of the magnetic field is in the toroidal direction.

Panel (d) shows the flow crosses the light cylinder radius at a cylindrical radius of $182~r_\ug$, the location where the blue line has a value of 1. Since the rotational velocity of the field lines is constant, this means the matter is moving along field lines that are bend backwards. In this solution $x_\uA = 0.1$, so the Alfv\'en point is crossed at a cylindrical radius of $18.2~r_\ug$.

\begin{figure*}
\begin{center}
\includegraphics[width = \textwidth]{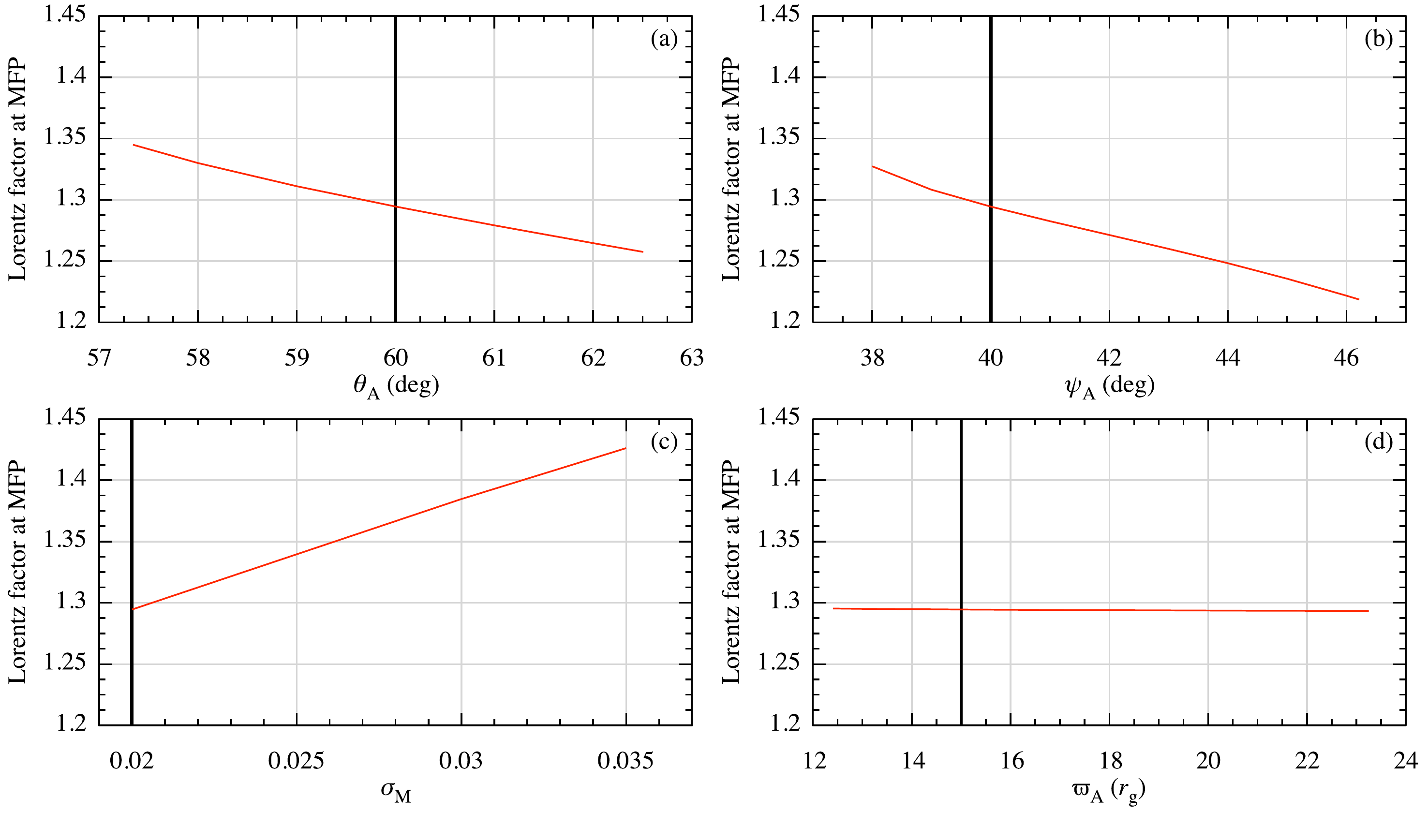}
\end{center}
\caption{
Lorentz factor at the MFP as a function of the four principal free parameters. The reference solution (see Table \ref{tab:fullgravityparametertable}) is indicated by the vertical black line. Parameters are as in Fig.~\ref{fig:fullMFPMSPlocation}. The vertical axis in each of these plots has the same range for comparison.
}
\label{fig:fullLorentz}
\end{figure*}

Although for this solution it was possible to integrate through the MSP and MFP, it is likely numerical artefacts developed at the crossings where the wind equation is close to 0/0, which would grow during the remainder of the integration. Therefore, we do not consider the integration below the MSP and beyond the MFP very reliable and we will use these two boundary conditions to match our solution to an inflow solution and a kinetically dominated jet, respectively.

\subsection{Exploring parameter space}
From the first solution, we changed fitting parameters to $x_\uA^2$ and $q$ to obtain a solution with round values for the remaining free parameters. The parameter values of this `reference' solution are given in Table \ref{tab:fullgravityparametertable}. From this solution, we increased and decreased each of the four free parameters in turn to determine the effect they have on the height of the MFP and the Lorentz factor at the MFP\@. The step sizes used are $1^\circ$ for $\theta_\uA$ and $\psi_\uA$, 0.01 for $\sigma_\uM$ and $1~r_\ug$ for $\varpi_\uA$, and these were decreased when no solution could be found in order to accurately map the boundaries of parameter space.

\begin{figure*}
\begin{center}
\includegraphics[width = 0.97 \textwidth]{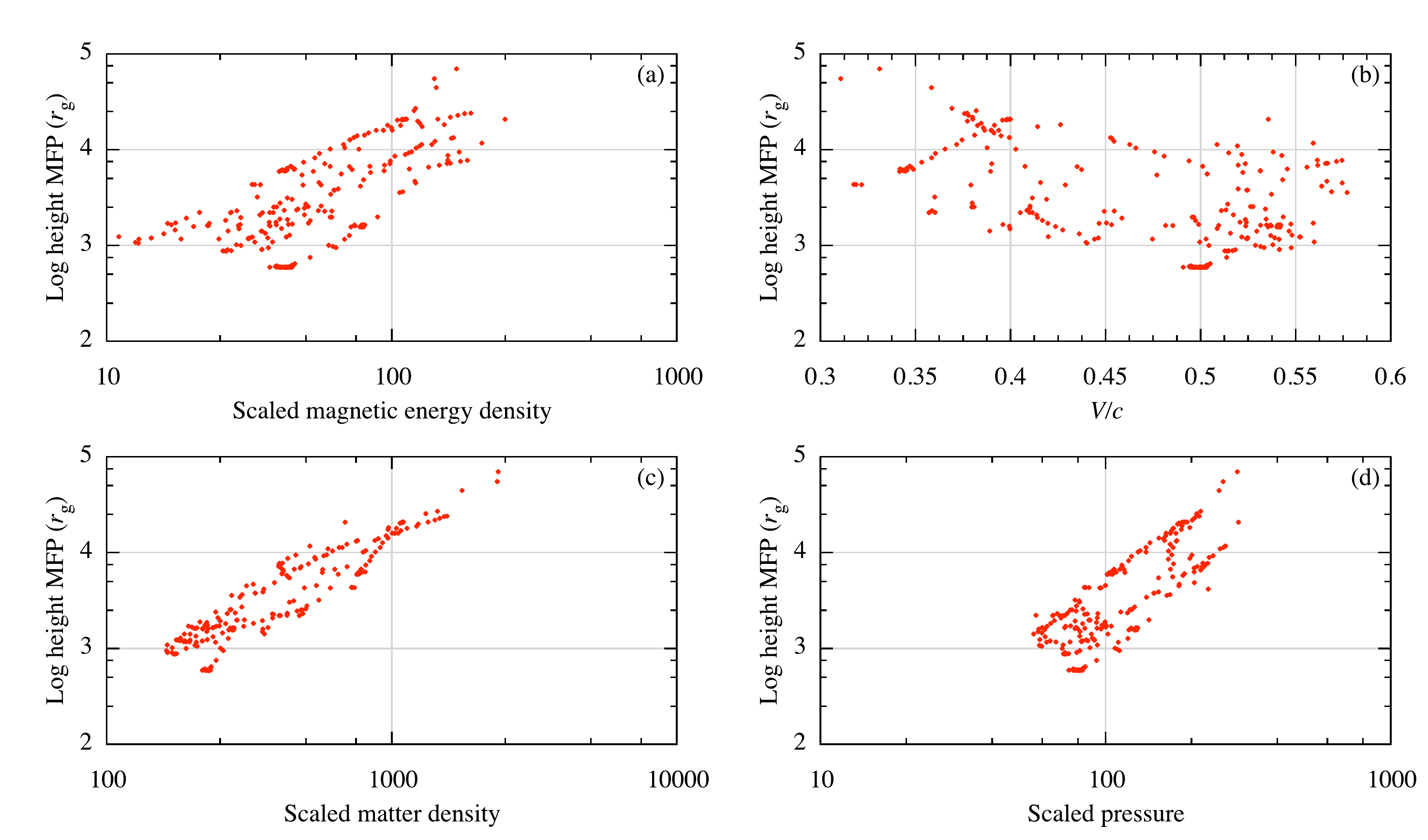}
\end{center}
\caption{
Heights of the MFP (in gravitational radii) as a function of the physical quantities at the MSP\@. Panel (a) shows the scaled magnetic energy density [$B^2 / (8 \uppi B_0^2 \alpha^{F - 2})$], panel (b) shows the velocity in units of the speed of light, panel (c) shows the scaled matter density [$4 \uppi \rho_0 / (B_0^2 \alpha^{F - 2})$] and panel (d) shows the scaled pressure [$4 \uppi P / (B_0^2 \alpha^{F - 2})$].
}
\label{fig:physMFPlocation}
\end{figure*}

\begin{figure*}
\begin{center}
\includegraphics[width = 0.97 \textwidth]{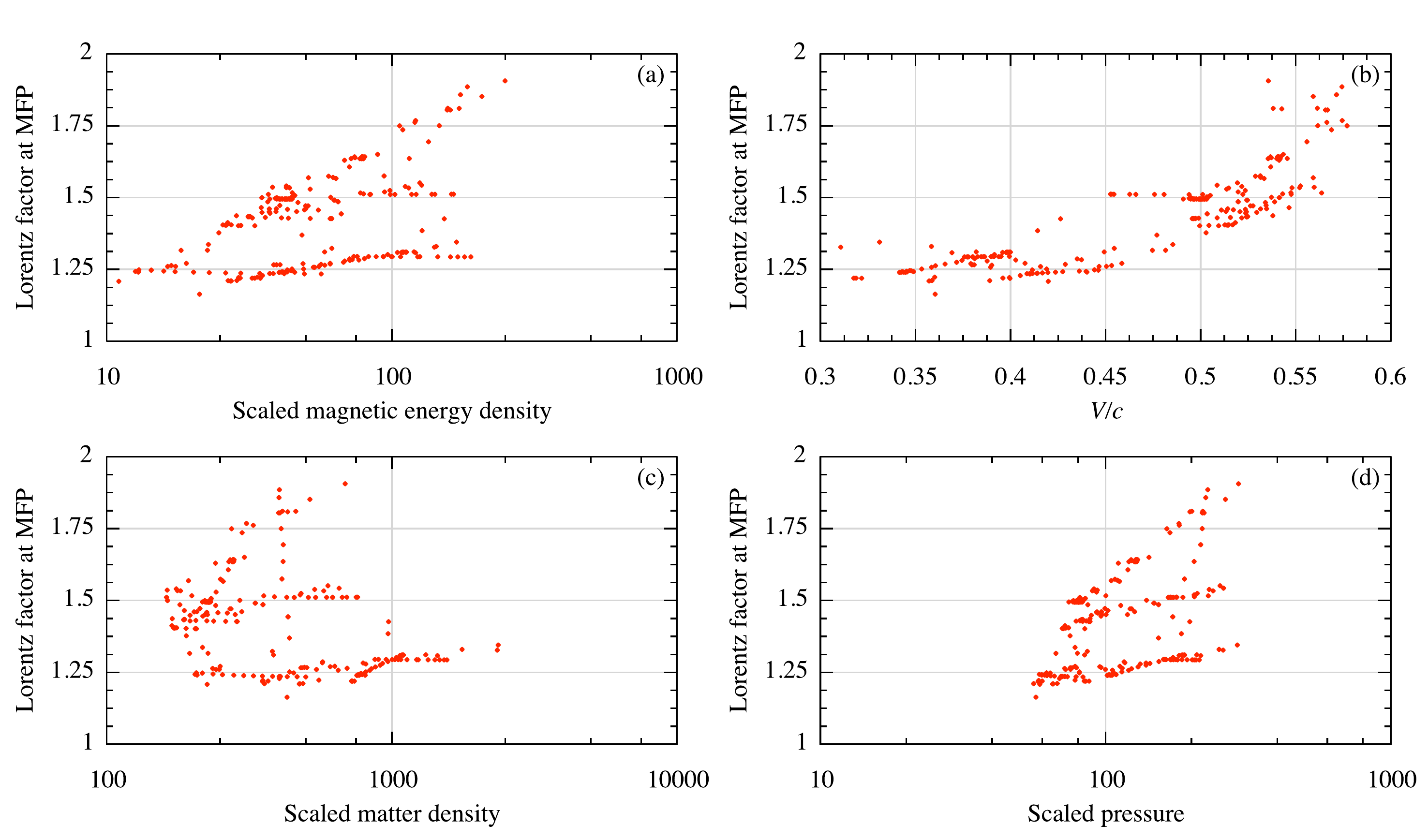}
\end{center}
\caption{
Lorentz factor at the MFP as a function of the physical quantities at the MSP\@. Physical quantities are as in Fig.~\ref{fig:physMFPlocation}.
}
\label{fig:physLorentz}
\end{figure*}

Fig.~\ref{fig:fullMFPMSPlocation} shows that the height of the MFP is strongly anticorrelated with the angle of the field line ($\psi_\uA$) and the angle of the Alfv\'en point ($\theta_\uA$), and weakly correlated with the cylindrical radius of the Alfv\'en point ($\varpi_\uA$). In other words, the height of the MFP decreases if the Alfv\'en point lies closer to the disc and the field lines there are more cylindrical. These trends correspond roughly to those found in Paper II. The range of magnetisation ($\sigma_\uM$) giving solutions is very small for these parameter values, so it is hard to make a general statement about its correlations. The height of the MSP is correlated with the height of the MFP for $\theta_\uA$ and $\psi_\uA$ and anticorrelated, albeit weakly, for $\sigma_\uM$ and $\varpi_\uA$. Since we are restricting ourselves to solutions that cross both the MSP and MFP, that the MSP approaches the Alfv\'en point for the lower values of $\theta_\uA$ and $\psi_\uA$ can be a possible reason for why it is impossible to find solutions below certain angles.

Fig.~\ref{fig:fullLorentz} shows the effect the parameters have on the value of the Lorentz factor at the MFP\@. Although the Lorentz factor can increase beyond this point, since we end our integration at the MFP, we have no information on its evolution there. As is already well known, the most pronounced effect on the Lorentz factor is the level of magnetisation, with a significant increase for only a small change in parameter value \citep{1992ApJ...394..459L}. Also $\theta_\uA$ and $\psi_\uA$ have a strong but anticorrelated effect, while the Lorentz factor is almost independent of the radius of the Alfv\'en point ($\varpi_\uA$; models with different $\varpi_\uA$, but otherwise the same parameters, are part of a quasi-self-similar family of models; see Section \ref{subsec:fullselfsimilarity}). The Lorentz factors for these solutions are rather small, lying in the range 1.22--1.43. This is mainly due to the small values of the magnetisation parameter.

\begin{figure*}
\begin{center}
\includegraphics[width = \textwidth]{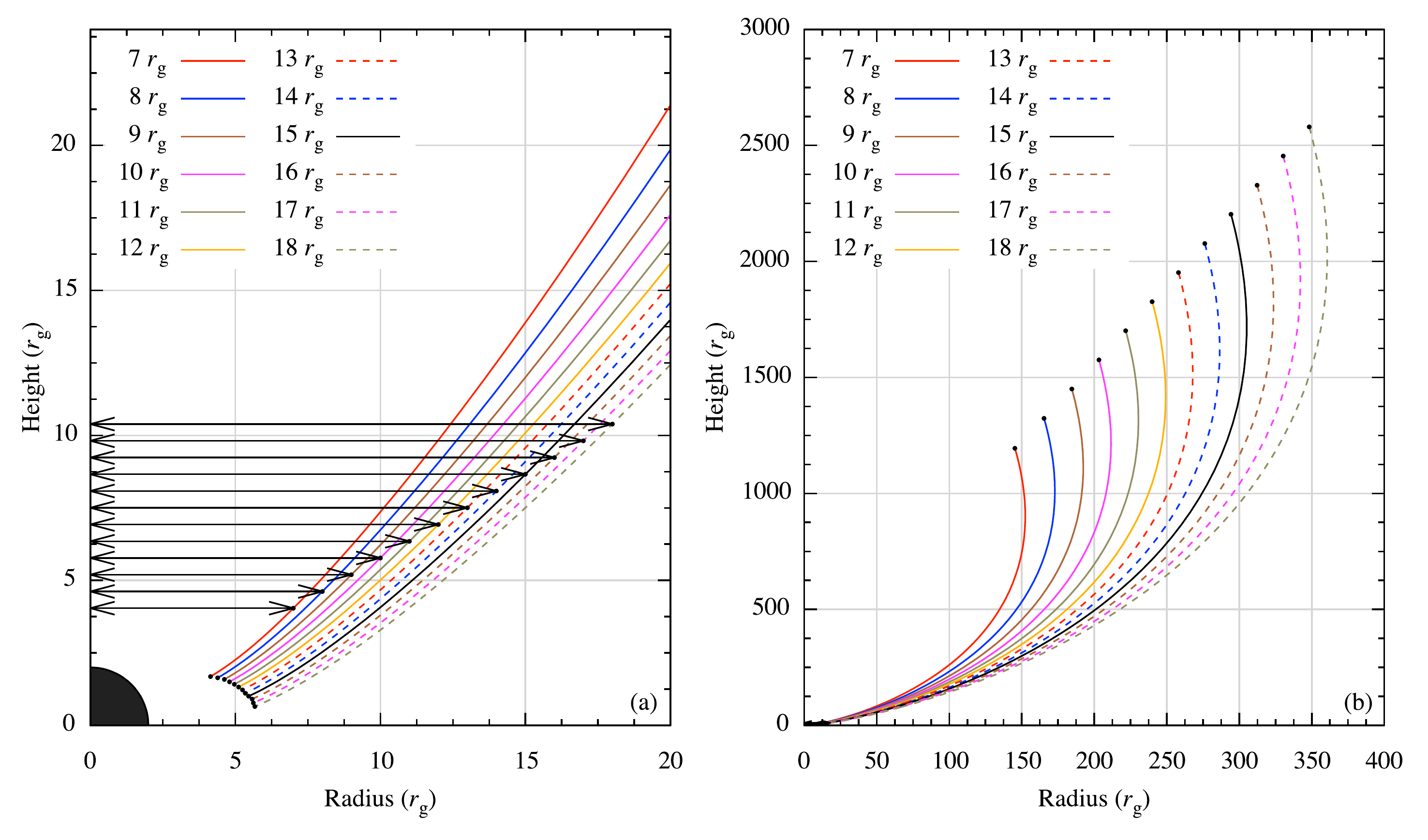}
\end{center}
\caption{
The effect of changing the cylindrical radius of the Alfv\'en point on the field line geometry. $\varpi_\uA$ ranges from 7--18~$r_\ug$ in steps of 1~$r_\ug$. In panel (a), the MSP and Alfv\'en point are indicated by the black dots. The arrows point to the Alfv\'en point. In panel (b) the MFP is indicated by black dots. Due to the extrapolation method, the MSP and MFP are also the end point of the field lines. Note that the height of the MSP (indicated at the lower end of the line) can be smaller than the Schwarzschild radius, but its spherical radius always remains outside the horizon. Note also that, while the Alfv\'en point and MFP positions are radially quasi-self-similar, those of the MSP are not.
}
\label{fig:fullalfvendistance}
\end{figure*}

Fig.~\ref{fig:physMFPlocation} shows a scatter plot of the height of the MFP as a function of the physical quantities at the MSP (which correspond to the four free parameters) for all solutions found so far. The physical quantities are the magnetic energy density, the velocity, the matter density and the pressure. Although these plots show by no means the complete picture, due to the limited number of solutions, there seems to be a trend for a higher MFP height with increasing magnetic energy density, matter density and pressure, although there is significant scatter. There seems to be no correlation between the velocity at the MSP and the height of the MFP\@.

Fig.~\ref{fig:physLorentz} shows a similar scatter plot for the Lorentz factor at the MFP as a function of the same quantities. Here the correlation with the velocity at the MSP is strong, while for the magnetic energy density, matter density and pressure there seems to be only a weak or no correlation. Based on these results, we expect sources with low MFP heights and high Lorentz factors to have a low magnetic energy density, matter density, and pressure close to the black hole, but high initial velocity.

If we want to find solutions with different properties, it is unlikely a single parameter will suit our needs. A compounding problem is that there is only a limited region in parameter space in which we can find solutions using the method outlined in Section \ref{subsec:approach}. Therefore, we cannot simply change a parameter to any value, as is clearly demonstrated by the limited range in magnetisation in Fig.~\ref{fig:fullLorentz}. Thus, in general, finding the desired solutions is a case of carefully navigating parameter space avoiding the boundaries.

\subsection{Self-similarity}
\label{subsec:fullselfsimilarity}
Including gravity into the relativistic equations inevitably violates the assumption of self-similarity, which means we cannot simply scale a single solution to populate a jet. However, by calculating many solutions with different Alfv\'en point radii, while keeping the other free parameters fixed, we can populate a jet that is quasi-self-similar. The reason for choosing $x_\uA^2$ and $q$ as fitting parameters, is that the field lines of the resulting solutions do not cross.

\begin{figure}
\begin{center}
\includegraphics[width = 0.47 \textwidth]{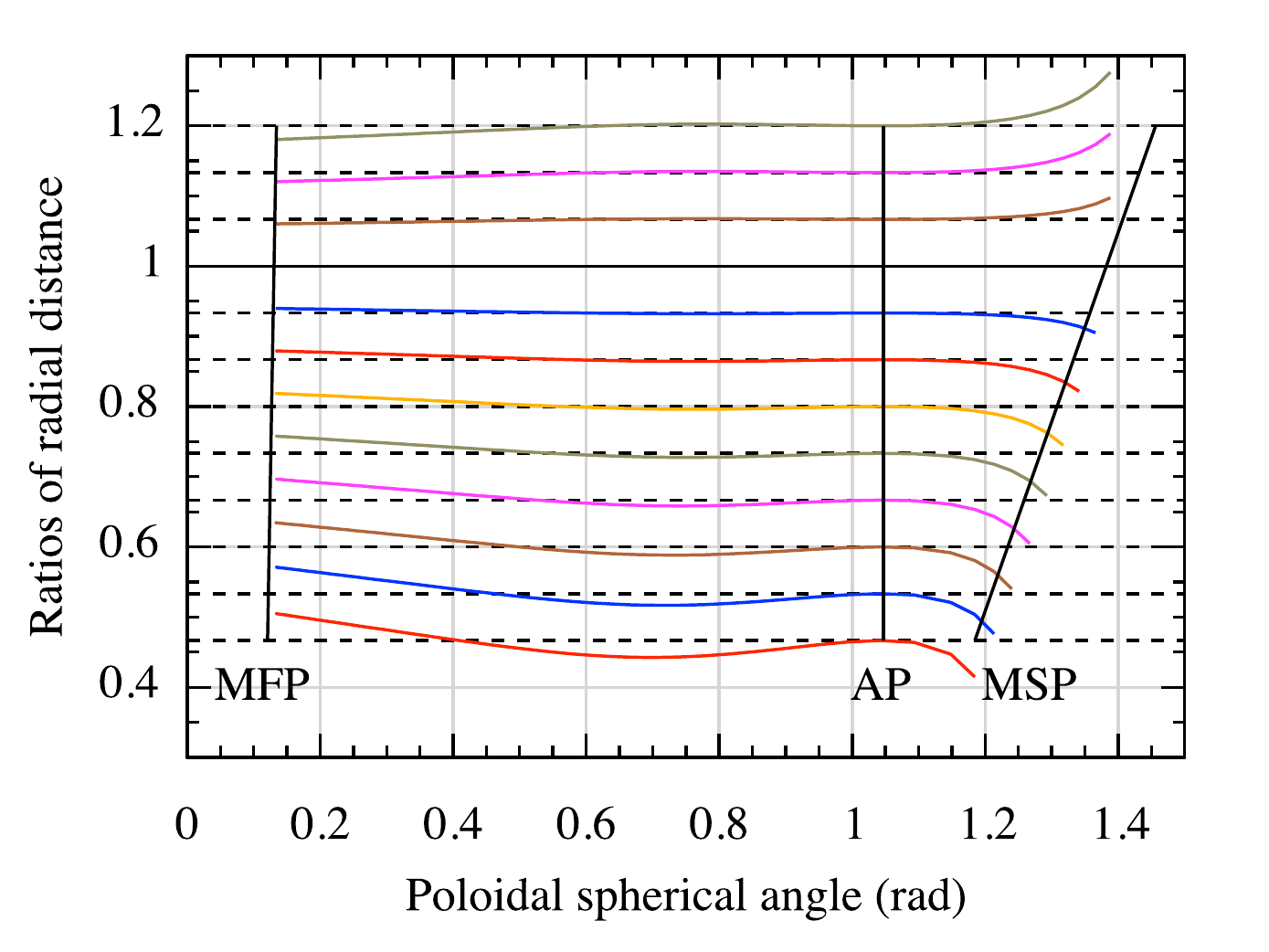}
\end{center}
\caption{
The ratios of the radial size of the field lines shown in Fig.~\ref{fig:fullalfvendistance} with regularly increasing $\varpi_\uA$ fitted with $x_\uA^2$ and $q$, colour-coded in the same way. The horizontal dashed lines show the values for exact self-similarity. The bottom line has $\alpha \approx 0.22$ and the top line has $\alpha = 1.44$ (where $\alpha \equiv \varpi_\uA^2/\varpi_0^2$), with respect to the reference field line, denoted by the horizontal black line. The parameters of the reference solution are $x_\uA^2 = 0.0902$, $\sigma_\uM = 0.02$, $q = 0.106$, $\varpi_\uA = 12~r_\ug$, $\mathcal{M} = 10~\text{M}_\odot$, $\theta_\uA = 60^\circ$, $\psi_\uA = 45^\circ$. The approximate locations of the MFP, Alfv\'en point (AP) and MSP are marked by the vertical lines.
}
\label{fig:fullselfsimilarity}
\end{figure}

In order to assess our solutions, we would like to know to what extent the jet provided by this method remains quasi-self-similar. Fig.~\ref{fig:fullalfvendistance} shows the poloidal projection of several field lines, both close to the black hole and far away. Fig.~\ref{fig:fullselfsimilarity} is the ratio of radii of these field lines with respect to the poloidal spherical angle. Perfectly self-similar field lines would be horizontal lines in this latter plot. Since the poloidal angle of the Alfv\'en point is fixed for all solutions, the Alfv\'en point is geometrically self-similar. The angle of the MFP is almost equal for all these solutions. The reason the ratios of the field lines at the MFP appear to be too small, is due to the fact we fix the ratios at the Alfv\'en point, even though exact self-similarity has not been established there due to the influence of gravity. The main deviation occurs around the MSP\@. Since the effects of gravity are the strongest there, this is expected. The lines with $\alpha > 1$ do not reach their MSPs because the reference solution, and therefore the ratio does not exist beyond its MSP\@. For the same reason, the line denoting the MSPs has been drawn based on the dashed lines and is only approximate. While the field lines start to deviate at the MSP, this deviation is just above 10 percent, so if we take that as an acceptable limit, the majority of these field lines are allowed within the flux tube.

Apart from geometry, the physical quantities such as the magnetic field strength, density and temperature also change from field line to field line, due to the changes in the fitting parameters. While in general $x_\uA^2$ only changes by a few percent, the changes in $q$ can be a factor of several. Since this change can significantly alter the temperature of the matter, limiting the deviations of these quantities from self-similarity can be an additional constraint on the size of the flux tube in which we consider our solutions.

\section{Discussion and conclusions}
\label{sec:fulldiscussion}
We have derived an expression for gravity that includes the kinetic, thermal and electromagnetic mass contributions (the `full' gravity term) from the general MHD equations in the special relativistic limit. The approach taken here is independent of the previous work to include gravity in the relativistic equations (the `kinetic' gravity term), but in the non-relativistic limit they are equivalent. This equivalence shows that the approach of the bridging solution introduced in Paper II is valid. The work presented here builds on that method by including the thermal and electromagnetic mass contributions, and taking the gravitational potential into account, which means that the current model is more physical, especially close to the black hole, and applicable to a wider range of boundary conditions.

If the gravitational potential energy is neglected, the energy-to-mass flux ratio ($\mu c^2$), the sum of the kinetic, thermal and electromagnetic energy, is a constant. Since we take the gravitational energy contribution into account, this quantity becomes a variable. The gravitational energy increases with height, and because the total energy is constant, $\mu$ has to decrease with height. Since the Lorentz factor is proportional to $\mu$, the counterintuitive result is that the Lorentz factor can decrease with height, even though the flow is thermally or magnetically accelerated (see Fig.~\ref{fig:fulloverview}). While the matter is accelerated along the field lines, it loses more energy as it climbs out of the potential well. Only when the gravitational potential has been mostly overcome, does the Lorentz factor increase again due to magnetic acceleration, with a corresponding decrease in Poynting flux.

The scaling of $(1 - x^2)$ in two parts of the full gravity term, including the previously derived kinetic mass contribution, is the result of the electric force in the transfield equation. Since the electric field is zero in VTST00, the scaling there reduces to 1. In the non-relativistic limit where the thermal and electromagnetic contributions are negligible, the full gravity term thus reduces to the kinetic gravity term.

We performed an initial parameter study in order to explore the trends in the solutions and showed that the results are broadly similar to those in the previous work. We have found solutions with MFP heights down to $600~r_\ug$ and with Lorentz factors below 2, due to the low magnetisation parameter. The boundaries in parameter space make a comprehensive exploration difficult.

There are several causes for these boundaries, limiting the region where solutions exist. The method we employ to find solutions, relies on the $x_\uA^2$,$q$-plane being regular, meaning it is possible to discern whether we need to increase or decrease a parameter in order to converge to an MSP or MFP\@. For certain parameters the used precision is not high enough to avoid numerical artefacts, showing many possible solutions scattered in a finite neighbourhood in the fitting plane. Since there is not a unique solution anymore, it is impossible to converge to the correct values. Also, for some unclear reason, in certain parts of parameter space the numerator or denominator can turn away from zero at either end of the integration, making it impossible to determine how to change the fitting parameters and find a solution. Another boundary may be caused by the coinciding of the MSP and Alfv\'en point.

The full gravity term has the same $\alpha$-dependence as the kinetic gravity term. Therefore, the region where gravity is unimportant and the region close to the black hole have a different self-similar scaling. By combining a varying number of solutions with different Alfv\'en point radii, we can construct flux tubes that satisfy self-similarity to any desired degree. The size of these flux tubes is limited by geometrical deviations and deviations in the physical quantities along different field lines. For this reason, we limit our solutions to a flux tube of a certain width, corresponding to the specified deviation of the field lines from self-similarity. The location of the start of particle acceleration seems to be relatively well defined in jets, while a wide jet has a large range of possible shock locations (see Fig.~\ref{fig:fullalfvendistance}). Combined with the limited depth of the jet photosphere, we believe a relatively narrow jet is a good enough approximation to calculate the expected emitted radiation. Therefore, we can use these flux tubes to relate the height of the MFP, and therefore the start of the acceleration region, to conditions very close to the black hole. This approach allows for a self-consistent determination of this location and thus of the conditions at the base of the jet from observed broad-band spectra of BHXRBs and AGN.
\vspace{-0.3cm}
\section*{Acknowledgments}
PP and SM gratefully acknowledge support from a Netherlands Organisation for Scientific Research (NWO) Vidi Fellowship. In addition, SM is grateful for support from the European Community's Seventh Framework Programme (FP7/2007-2013) under grant agreement number ITN 215212 `Black Hole Universe'. Part of this research was carried out at the Jet Propulsion Laboratory, California Institute of Technology, under contract with the National Aeronautics and Space Administration. We thank the anonymous referee for helpful comments that improved this manuscript.

\appendix
\section{Equations}
\label{app:equations}
The energy equation is given by
\begin{IEEEeqnarray}{rCl}
\IEEEeqnarraymulticol{3}{l}{\frac{\mu^2}{\xi^2} \frac{G^4 (1 - M^2 - x_\uA^2)^2 - x^2 (G^2 - M^2 - x^2)^2}{G^4 (1 - M^2 - x^2)^2}} \IEEEnonumber \\
\qquad & = & 1 + \frac{F^2 \sigma_\uM^2 M^4 \sin^2(\theta)}{\xi^2 x^4 \cos^2(\psi + \theta)}.
\label{eq:energyequation}
\end{IEEEeqnarray}
If gravity is included, due to the gravitational potential energy $\mu$ is no longer a constant of motion along the field line, as described in Section \ref{subsec:gravityterm}.

If we multiply the terms $A_1$ through $C_1$ by a factor, this factor cancels in the wind equation, so it is possible to use an arbitrary scaling. The same is true for the terms $A_2$ through $C_2$. These latter terms have a scaling based on equations (A8), not (B2e), in \citet{2003ApJ...596.1080V}, which is why they differ from the terms given in Paper I and II. Without the full gravity term they reduce to the same result, but with full gravity certain terms in the equations do not cancel, and the equations given in this paper should be used instead. They therefore represent a new result for this paper. The equations given here do not contain gravity themselves: the terms $C_1^+$ and $C_2^+$ should be added to $C_1$ and $C_2$, respectively, before multiplication with $B_2$ and $B_1$ in equation \eqref{eq:windequation}. The easiest form to cast $A_1$ through $C_2$ in are
\begin{IEEEeqnarray}{rCl}
A_1 & = & \Bigg( \frac{\mu^2 x_\uA^6}{F^2 \sigma_\uM^2} \frac{M^2}{G^2} \frac{(1 - G^2)^2}{(1 - M^2 - x^2)^3} \frac{\cos^3(\psi + \theta)}{\sin^2(\theta) \sin(\psi + \theta)} \IEEEnonumber \\
& & - \frac{x_\uA^4}{F^2 \sigma_\uM^2} \frac{\xi^2}{M^2} \frac{(\Gamma - 1) (\xi - 1)}{\left( 2 - \Gamma \right) \xi + \Gamma - 1} \frac{\cos^3(\psi + \theta)}{\sin^2(\theta) \sin(\psi + \theta)} \IEEEnonumber \\
& & + \frac{M^2}{G^4} \frac{\cos(\psi + \theta)}{\sin(\psi + \theta)} \Bigg), \\
B_1 & = & \frac{M^4}{G^4}, \\
C_1 & = & \frac{\xi^2 x_\uA^4}{F^2 \sigma_\uM^2} \frac{\cos(\psi) \cos^2(\psi + \theta)}{\sin^3(\theta) \sin(\psi + \theta)} \bigg\{ \frac{\mu^2}{\xi^2} \frac{G^4 (1 - M^2 - x_\uA^2)^2}{G^4 (1 - M^2 - x^2)^2} - 1 \IEEEnonumber \\
& & + 2 x^2 \bigg[ \frac{\mu^2}{\xi^2} \frac{G^4 (1 - M^2 - x_\uA^2)^2 - x^2 (G^2 - M^2 - x^2)^2}{G^4 (1 - M^2 - x^2)^3} \IEEEnonumber \\
& & - \frac{\mu^2}{\xi^2} \frac{G^2 (G^2 - M^2 - x^2) (1 - x_\uA^2)}{G^4 (1 - M^2 - x^2)^2} \bigg] \bigg\}, \\
A_2 & = & \frac{\xi^2 x_\uA^4}{F^2 \sigma_\uM^2} \sin(\psi + \theta) \cos(\psi + \theta) \IEEEnonumber \\
& & \times \bigg[ \frac{\mu^2}{\xi^2} \frac{x^2 (1 - G^2)^2}{(1 - M^2 - x^2)^3} - \frac{(\Gamma - 1) (\xi - 1)}{\xi - (\Gamma - 1) (\xi - 1)} \frac{G^4}{M^4} \bigg], \\
B_2 & = & \left[ \frac{(1 - x^2)}{\cos^2(\psi + \theta)} - M^2 \right] \sin^2(\theta), \\
C_2 & = & \frac{G^4}{M^2} \frac{\xi^2 x_\uA^4}{F^2 \sigma_\uM^2} \frac{\cos(\psi) \sin(\psi + \theta)}{\sin(\theta)} \bigg\{ \frac{\mu^2}{\xi^2} \frac{G^4 (1 - M^2 - x_\uA^2)^2}{G^4 (1 - M^2 - x^2)^2} - 1 \IEEEnonumber \\
& & + \frac{2 x^2}{1 - M^2 - x^2} \left[ \frac{\mu^2}{\xi^2} \frac{G^2 M^2 (1 - G^2) (1 - M^2 - x_\uA^2)}{G^4 (1 - M^2 - x^2)^2} \right] \bigg\} \IEEEnonumber \\
& & + 2 \frac{\Gamma - 1}{\Gamma} \frac{F - 2}{F^2 \sigma_\uM^2} \frac{\xi (\xi - 1) x^4}{M^2} + 2 x^2 \frac{\cos(\psi) \sin(\theta) \sin(\psi + \theta)}{\cos^2(\psi + \theta)} \IEEEnonumber \\
& & + (1 - M^2 - x^2) \frac{\cos(\psi) \sin(\theta) \sin(\psi + \theta)}{\cos^2(\psi + \theta)} \IEEEnonumber
\end{IEEEeqnarray}
\begin{IEEEeqnarray}{rCl}
& & + \frac{\sin^2(\theta)}{\cos^2(\psi + \theta)} \left[ F - 2 - F x^2 + x^2 \right] + \frac{x_\uA^4 \mu^2 x^2}{F^2 \sigma_\uM^2} \IEEEnonumber \\
& & \times \left[ \left( \frac{1 - G^2}{1 - M^2 - x^2} \right)^2 (F - 1) - \frac{1}{M^2} \left( \frac{G^2 - M^2 - x^2}{1 - M^2 - x^2} \right)^2 \right]. \IEEEnonumber \\
\end{IEEEeqnarray}

The transfield equation \emph{without} gravity can be reconstructed from $A_2$, $B_2$ and $C_2$ by using equation \eqref{eq:determinant}.

\label{lastpage}

\bibliographystyle{mn2e}
\bibliography{references}

\end{document}